%% file: techreport_final.tex
\documentclass{article}
\usepackage{times}
\usepackage{graphicx}
\usepackage{amsfonts}
\usepackage{amssymb}
\usepackage{amsmath} 
\usepackage{amsthm}
\usepackage{rotating}

\usepackage{natbib}
\usepackage{graphicx}

\usepackage{algorithm}
\usepackage{algorithmic}

\usepackage[margin=1in]{geometry}

\newtheorem{proposition}{Proposition}
\newtheorem{definition}{Definition}

\newcommand\one[1]{{\bf 1}\left(#1\right)}

\def\inputSet{\mathcal{S}}
\def\inputSpace{\mathcal{X}}

\def\realSet{\mathbb{R}}
\def\naturalSet{\mathbb{N}}
\def\sign{\text{sign}}
\def\featureSpace{\mathcal{H}}

\input{header.tex}

\title{The pharmacophore kernel for virtual screening with support vector machines}
\author{
Pierre Mah{\'e}\\
Center for Computational Biology\\
Ecole des Mines de Paris\\ 
\texttt{Pierre.Mahe@ensmp.fr} 
\and 
Liva Ralaivola\\
Laboratoire d'Informatique Fondamentale\\
University Provence/Aix-Marseille 1\\ 
\texttt{Liva.Ralaivola@cmi.univ-mrs.fr}
\and 
V{\'e}ronique Stoven\\
Center for Computational Biology\\
Ecole des Mines de Paris\\ 
\texttt{Veronique.Stoven@ensmp.fr}
\and 
Jean-Philippe Vert\\
Center for Computational Biology\\
Ecole des Mines de Paris\\ 
\texttt{Jean-Philippe.Vert@ensmp.fr}
}
\begin{document}

\maketitle

\begin{abstract}
We introduce a family of positive definite kernels specifically optimized for the manipulation of 3D structures of molecules with kernel methods. The kernels are based on the comparison of the three-points pharmacophores present in the 3D structures of molecules, a set of molecular features known to be particularly relevant for virtual screening applications. We present a computationally demanding exact implementation of these kernels, as well as fast approximations related to the classical fingerprint-based approaches. Experimental results suggest that this new approach outperforms state-of-the-art algorithms based on the 2D structure of molecules for the detection of inhibitors of several drug targets.
\end{abstract}

\section{Introduction}
Virtual screening refers to the process of inferring biological properties of molecules \emph{in silico}, and plays an increasingly important role at the early stages of the drug discovery process to select candidate molecules with promising drug-likeness, including good toxicity and pharmacokinetics properties, as well as the potential to bind and inhibit a target protein of interest \cite{Manly2001impact}. In this context, structure-activity relationship (SAR) analysis is commonly used to build predictive models for the property of interest from a description of the molecules, using statistical procedures to build these models from the analysis of molecules with known properties \cite{Butina2002Predicting}.

It is widely accepted that several drug-like properties can be efficiently deduced from the 2D  structure of the molecule, that is, the description of a molecule as a set of atoms and their covalent bonds. For example, Lipinski's ``rule of five'' remains a widely used standard for the prediction of intestinal absorption \cite{Lipinski2001Experimental}, and the prediction of mutagenicity from 2D molecular fragments is an accurate state-of-the-art approach \cite{King1996Structure-activity}. In the case of target binding prediction, however, the molecular mechanisms responsible for the binding are known to depend on a precise 3D complementarity between the drug and the target, from both the steric and electrostatic perspectives \cite{Bohm2003Protein-ligand}. For this reason, there has been a long history of research on the prediction of these interactions from the 3D representation of molecules, that is, their spatial conformation in the 3D space. If the 3D structure of the target is known, the strength of the interaction can be directly evaluated by docking techniques, that quantify the complementarity of the molecule to the target in terms of energy \cite{Halperin2002Principles}. In the general case where the 3D structure of the target is unknown, however, the docking approach is not possible anymore and the modeler must resort to creating a predictive model from available data, typically a pool of molecules with known affinity to the target; this approach is usually referred to as the \emph{ligand-based} approach to virtual screening.

Most approaches to ligand-based virtual screening require to represent and compare 3D structures of molecules. The comparison of 3D structures can for example rely on optimal alignments in the 3D space \cite{Lemmen2000Computational}, or on the comparison of features extracted from the structures \cite{Xue2000Molecular}. Features of particular importance in this context are subsets of two to four atoms together with their relative spatial organization, also called \emph{pharmacophores}. Discovering pharmacophores common to a set of known inhibitors to a drug target can be a powerful approach to the screening of other candidate molecules containing the pharmacophores, as well as a first step towards the understanding of the biological phenomenon involved \cite{Holliday1997Using,Finn1998Pharmacophore}. Alternatively, pharmacophore fingerprints, that is, bitstrings representing a molecule by the pharmacophores it contains, has emerged as a potential approach to apply statistical learning methods for SAR, although sometimes with mixed results \cite{Matter1999Comparing,Brown1997information,Bajorath2001Selected}.

We focus in this paper on an extension of the fingerprint representation of molecules for building SAR models with support vector machines (SVM). SVM is an algorithm for learning a classification or regression rule from labeled examples \cite{Boser1992training,Vapnik1998Statistical}, that has recently been subject to much investigations for SAR applications in chemoinformatics \cite{Burbidge2001Drug,Byvatov2003Comparison,Mahe2005Graph,Ralaivola2005Graph}. Although SVM can be trained from a vector or bitstring representation of molecules, they can also take advantage of a mathematical trick to only rely on a measure of similarity between molecules, known as \emph{kernel}. This trick, common to other algorithms called kernel methods \cite{Scholkopf2002Learning}, was for example used in \cite{Kashima2004Kernels,Mahe2005Graph} to build SAR models from a 2D fingerprint of molecules of virtually infinite length. Here we investigate the possibility to use this trick in the context of 3D SAR modeling. We propose a measure of similarity between 3D structures, which we call the \emph{pharmacophore kernel}, based on the comparison of pharmacophores present in the structures. It satisfies the mathematical properties required to be a valid kernel and it therefore allows the use of SVM for model building. This kernel bears some similarity with pharmacophore fingerprint approaches, although it produces more general models. In fact, we show that a fast approximation of this kernel, based on pharmacophore fingerprints, leads to significantly lower performance on a benchmark dataset. We also show that competitive performance can be obtained by a fast fingerprint-based approximation using the Tanimoto coefficient as a valid kernel for SVM \cite{Ralaivola2005Graph}. The overall good performance of the approach on this benchmark supports its relevance as a potentially effective tool for 3D SAR modeling.

This paper is organized as follows. A light introduction to SVM and kernel methods is provided in Section \ref{sec:kernel-methods}, followed by the definition of the pharmacophore kernel (Section \ref{sec:pharma-kernel}). The exact computation of this kernel is presented in Section \ref{sec:computation}, followed by a discussion about the connection between the pharmacophore kernel and recently introduced graph kernels (Section \ref{sec:graph-kernels}) and the presentation of a fast approximation (Section \ref{sec:fast-approximation}). Experimental results on a benchmark dataset for inhibitor prediction are then presented in Section \ref{sec:experiments}, followed by a short discussion.

\section{Support vector machines}\label{sec:kernel-methods}
\begin{figure*}[tb]
\centering
\includegraphics[width=\textwidth]{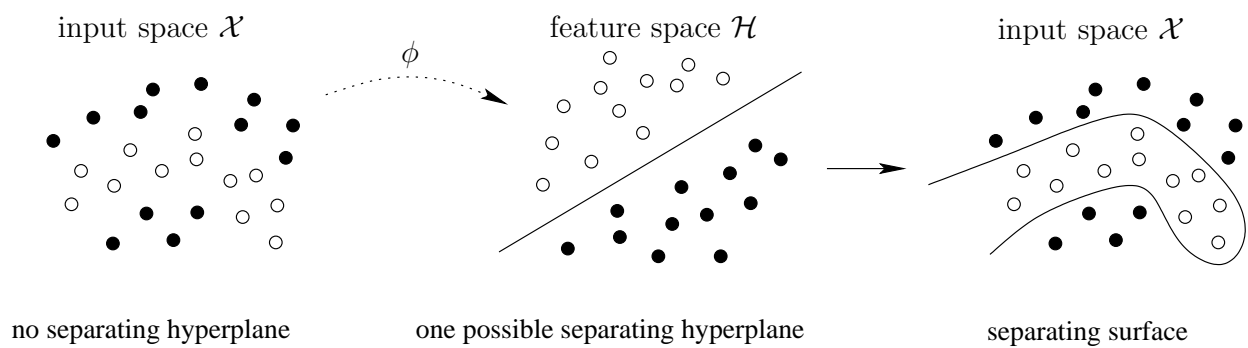}
\caption{The kernel trick: instead of looking for, e.g.,  a separating hyperplane directly in the input space $\inputSpace$,
training patterns (white and black disks) are mapped into a {\it feature space} $\featureSpace$ through a function $\phi$ in which 
a hyperplane is computed; this hyperplane might correspond to a complex surface in the input space. Using a proper
positive definite kernel $k$ when carrying out the computations to derive the separating hyperplane is equivalent to 
directly working with the images of the training samples by some mapping $\phi$ from $\inputSpace$ to some feature space $\featureSpace$;
the existence of $\phi:\inputSpace\rightarrow\featureSpace$ is guaranteed by (\ref{eqn:mercer}).
\label{fig:featurespace}}
\end{figure*}

In this section we briefly review the basics of support vector machines \cite{Boser1992training,Vapnik1998Statistical}. The interested reader is invited to refer to \cite{Burges1998Tutorial,Scholkopf2002Learning,Shawe-Taylor2004Kernel} for further details. In its simplest form SVM is an algorithm to learn a binary classification rule from a set of labeled examples. More formally, suppose one is given a set of examples with a binary label attached to each example, that is, a set $\inputSet=\{(x_1,y_1),\ldots,(x_\ell,y_\ell)\}$ where $(x_i,y_i)\in \inputSpace\times\{-1,+1\}$ for $i=1,\ldots,\ell$. Here $\inputSpace$ is an inner-product space (e.g. $\realSet^d$), equipped with inner product $\langle\cdot,\cdot\rangle$, that represents the space of data to be analyzed, typically molecules represented by $d$-dimensional fingerprints, and the labels $+1$ and $-1$ are meant to represent two classes of objects, such as inhibitors or non-inhibitors of a target of interest. The purpose of SVM is to learn from $\inputSet$ a classification function $f:\inputSpace\rightarrow\{-1,+1\}$ that can be used to predict the class of new unlabeled examples $x$ as $f(x)$.

In the case of SVM, the classification function is simply of the form $f(x)=\sign(\langle w,x\rangle+b)$, where $\sign(\cdot)$ is the function returning the sign, $+1$ or $-1$, of its argument. Geometrically speaking, this means that  $f$ outputs a prediction for a pattern $x$ depending upon which side of the hyperplane $\langle w,x\rangle+b=0$ it falls in. More precisely, SVM learn a separating hyperplane from $\inputSet$ defined by a vector $w$ that is a linear combination of the training vectors $w=\sum_{i=1}^\ell\alpha_ix_i$, for some
$\alpha_i\in\realSet,i=1,\ldots,\ell$, obtained by solving a linearly constrained quadratic problem meant to optimize a trade-off between finding a hyperplane that correctly separates all the points, while being as far as possible from each point. The linear classifier $f$ can consequently be rewritten as
\begin{equation}
f(x)=\sign\left(\sum_{i=1}^\ell\alpha_i\langle x_i,x\rangle+b\right)\;.
\label{eq:output}
\end{equation}

However, when dealing with nonlinearly separable problems, such as the
one depicted on Figure~\ref{fig:featurespace} (left), the set of
linear classifiers may not be rich enough to provide a good
classification function, no matter what the values of the parameters
$w\in\inputSpace$ and $b\in\realSet$ are. The purpose of the {\em
  kernel trick} \cite{Aizerman1964Theoretical,Boser1992training}, is
precisely to overcome this limitation by applying a linear approach to
the transformed data $\phi({x}_1),\ldots,\phi({x}_\ell)$ rather
than the original data, where $\phi$ is an embedding from
the input space $\inputSpace$ to the {\em feature space}
$\featureSpace$, usually, but not necessarily, a high-dimensional space,
equipped with dot product $\langle\cdot,\cdot\rangle_{\featureSpace}$.
Thus, according to~(\ref{eq:output}), the separating function $f$ writes as 
\begin{equation}
f(x)=\sign\left(\sum_{i=1}^\ell\alpha_i\langle \phi(x_i),\phi(x)\rangle_{\featureSpace}+b\right)\;.
\label{eq:output1}
\end{equation} 
The key ingredient in the kernel approach is to replace the dot 
product in $\featureSpace$ with a kernel, using the definition of 
positive definite kernels.

\begin{definition}[Positive definite kernel]
  Let $\inputSpace$ be a nonempty space.  Let
  $K:\inputSpace\times\inputSpace\rightarrow\realSet$ be a symmetric function. 
$K$ is said to be a positive
  definite kernel if and only if, for all $\ell\in\naturalSet$, for all
  $x_1,\ldots,x_\ell\in\inputSpace$, the square $\ell\times\ell$
  matrix $K=(K(x_i,x_j))_{1\leq i,j\leq\ell}$ is positive
  semi-definite, that is, all its eigenvalues are nonnegative.
\label{def:pdkernel}
\end{definition}
  
  For a given set $\inputSet_x=\{x_1,\ldots,x_\ell\}$, $K$ is
  the  {\em Gram matrix} of $K$ with respect to $\inputSet_x$. A fundamental property of positive definite kernels that underlies the kernel trick is the fact that each such kernel can be represented as an inner product in some space. More precisely, it can be shown \cite{Aronszajn1950Theory} that 
for any positive definite kernel function
  $K$, there exists a space
  $\featureSpace$, equipped with the inner product
  $\langle\cdot,\cdot\rangle_\featureSpace$, and a mapping
  $\phi:\inputSpace\rightarrow\featureSpace$ such that:
\begin{equation}\label{eqn:mercer}
\forall u,v\in\inputSpace\quad K(u,v)=\langle\phi(u),\phi(v)\rangle_\featureSpace\;.
\end{equation}
The kernel trick consists in replacing all occurrences of
$\langle\cdot,\cdot\rangle_\featureSpace$ in~(\ref{eq:output1}) by a positive definite kernel $K$ such that the
corresponding decision function $f$, for an input pattern $x$, is
given by:
\begin{equation}
f({x})=\sign\left(\sum_{i=1}^\ell\alpha_iK(x_i,x)+b\right)\;.
\label{eq:kerneloutput}
\end{equation}
For SVM as well as for other kernel methods, the knowledge of the Gram matrix suffices to
obtain the coefficients $\alpha_i$.

For any given positive definite kernel, applying the kernel trick turns out to be
equivalent to transforming the input patterns
$x_1,\ldots,x_\ell$ into the corresponding vectors
$\phi(x_1),\ldots,\phi(x_\ell)\in\featureSpace$ and to look for hyperplanes in
$\featureSpace$, as illustrated in Figure~\ref{fig:featurespace}
(middle). The decision surface in input space $\inputSpace$
corresponding to the selected separating hyperplane in $\featureSpace$
might be quite complex (see Figure~\ref{fig:featurespace}, right).

A noteworthy feature of support vector machines and more generally of kernel methods \cite{Shawe-Taylor2004Kernel} is that, since ready-to-use
libraries to derive separating hyperplanes are available, the only
requirement for them to be applied to a specific classification
problem is to have at hand a proper kernel function to assess the
similarity between patterns of the input space considered. Henceforth,
their use actually fit in the framework of classification problems
involving structured data such as chemical compounds, provided some
kernel function has been derived. The rest of the paper is devoted to the construction and analysis of such a kernel for 3D structures of molecules.

\section{The pharmacophore kernel}\label{sec:pharma-kernel}

\begin{figure*}
\centering
\includegraphics[width=0.24\textwidth]{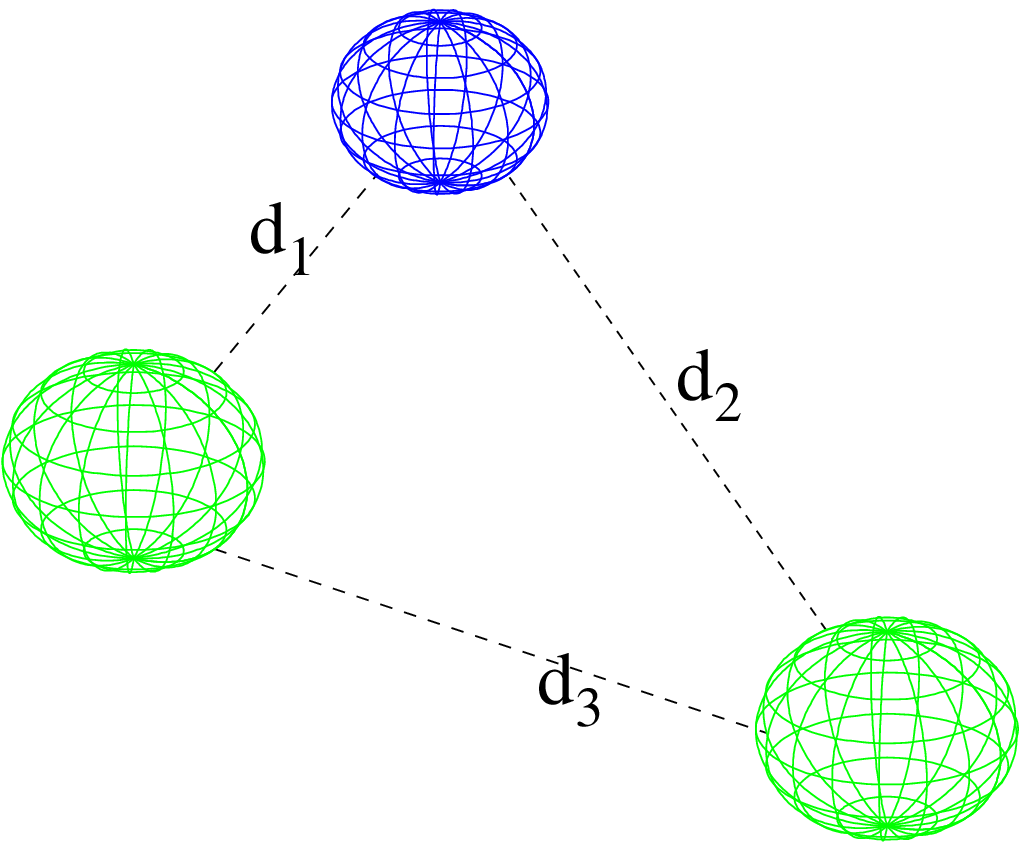}
\hspace{0.06\textwidth}
\includegraphics[width=0.24\textwidth]{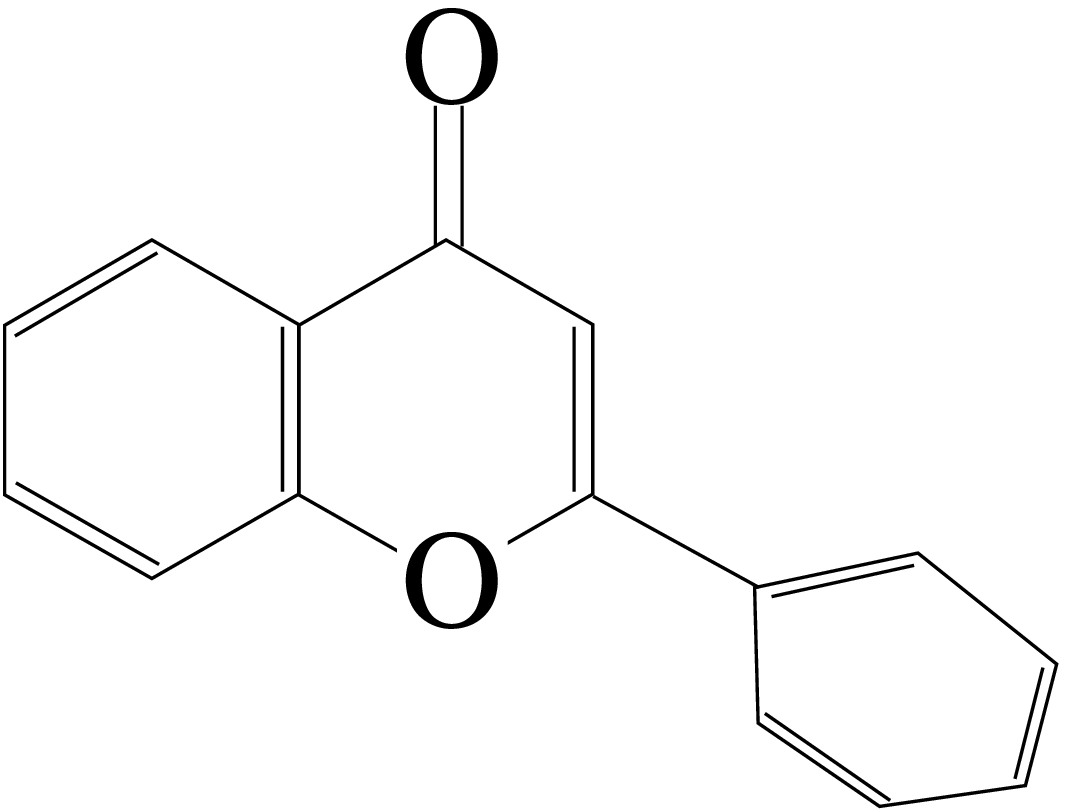}
\hspace{0.06\textwidth}
\includegraphics[width=0.24\textwidth]{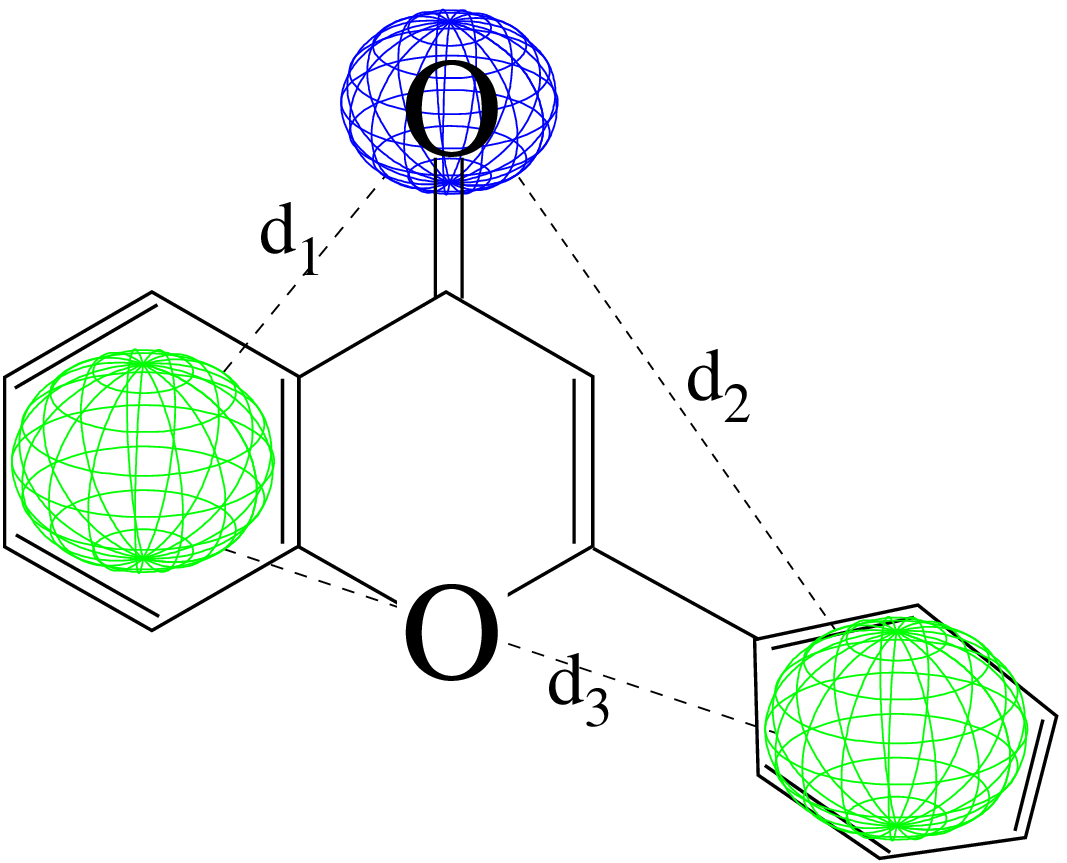}
\caption{Left: a 3-points pharmacophore made of one hydrogen bond acceptor (topmost sphere) and two aromatic rings, with distances $d_1$, $d_2$ and $d_3$ between the features. Middle: the molecule of flavone. Right: match between flavone and the pharmacophore. \label{fig:pharmacophore}}
\end{figure*}

A {\it pharmacophore} is usually defined as a three-dimensional arrangement of atoms - or groups of atoms - responsible for the biological activity of a drug molecule \cite{Guner2000Pharmacophore}. The present work focuses on \emph{three-points pharmacophores}, composed of three atoms whose arrangement therefore forms a triangle in the 3D space (Figure~\ref{fig:pharmacophore}). With a slight abuse we refer as pharmacophore below to \emph{any} possible configuration of three atoms or classes of atoms arranged as a triangle and present in a molecule, representing therefore a \emph{putative} configuration responsible for the biological property of interest. 

Throughout this paper we represent the 3D structure of a molecule as a set of points in $\mathbb{R}^3$. These points correspond to the 3D coordinates of the atoms of the molecule (for a given arbitrary basis of the 3D Euclidean space), and they are labeled with some information related to the atoms. More formally, we define a molecule $m$ as 
$$
m = \cbr{ \br{x_i,l_i} \in \RR^3 \times \mc{L} }_{i=1,...,\abs{m}} \; ,
$$ where $\abs{m}$ is the number of atoms that compose the molecule and $\mc{L}$ denotes the set of atom labels. The label is meant to contain the relevant information to characterize a pharmacophore based on atoms, such as the type of atom (C, N, O, ...) and its partial charge. The three-points pharmacophores considered in this work correspond to triplets of distinct atoms of the molecules. The set of pharmacophores of the molecule $m$ can therefore be formally defined as:
$$
\mc{P}(m) = \cbr{ \br{p_1,p_2,p_3} \in m^3, p_1 \neq p_2 \neq p_3 } \; .
$$
More generally, the set of all possible pharmacophores is naturally defined as $\mc{P} = \br{\RR^3 \times \mc{L}}^3$, to ensure the inclusion $\mc{P}(m) \subset \mc{P}$. We can now define a general family of kernels for molecules based on their pharmacophore content:
\begin{definition}\label{def:kernel}
For any positive definite kernel for pharmacophores $K_\mc{P}:\mc{P}\times \mc{P} \rightarrow \RR$, we define a corresponding \emph{pharmacophore kernel} for any pair of molecules $m$ and $m'$ by : 
\begin{equation}\label{eq:kernel}
K(m,m') := \sum_{ p \in \mathcal{P}(m)}  \sum_{ p' \in \mathcal{P}(m') } K_{\mathcal{P}} (p,p')\;,
\end{equation}
with the convention that $K(m,m') = 0$  if  either $\mathcal{P}(m)$ or $\mathcal{P}(m')$ is empty.
\end{definition}
The fact that the pharmacophore kernel defined in (\ref{eq:kernel}) is a valid positive definite kernel on the set of molecules, as soon as $K_{\mathcal{P}}$ is itself a valid positive definite kernel on the set of pharmacophores, is a classical result (see, e.g., \cite[Lemma 1]{Haussler1999Convolution}). The problem of constructing a pharmacophore kernel for molecules therefore boils down to the simpler problem of defining a kernel between pharmacophores.
A chemically relevant measure of similarity between pharmacophores should obviously quantify at least two features: first, similar pharmacophores should be made of similar atoms, and second, the atoms should have similar relative positions in the 3D space. It is therefore natural to study kernels for pharmacophores that decompose as follows:   
\begin{equation}\label{eq:kernel_pharma}
K_{\mathcal{P}}(p,p') = K_I(p,p') \times K_S(p,p') \;,
\end{equation}
where $K_I$ is a kernel function assessing the similarity between the triplets of basis atoms of the pharmacophores (their so-called \emph{intrinsic} similarity), and $K_S$ is a kernel function introduced to quantify their \emph{spatial} similarity.

We can furthermore investigate intrinsic and spatial kernels that factorize themselves as products of more basic kernels between atoms and pairwise distances, respectively. Triplets of atoms are indeed globally similar if the three corresponding pairs of atoms are simultaneously similar, and triangles are similar if the lengths of their edges are pairwise similar.
For any pair of pharmacophores $p = \br{\br{x_{i},l_{i}}}_{i=1,2,3}$ and $p' = \br{\br{x'_{i},l'_{i}}}_{i=1,2,3}$, this suggests to define kernels as follows:
\begin{eqnarray}
K_I(p,p') & = & \prod_{i=1}^3 K_{Feat} \br{ l_{i},l'_{i} }\label{eq:kernel_node} \; ,\\
K_S(p,p') & = & \prod_{i=1}^3 K_{Dist} \br{ \nm{x_i - x_{i+1}} , \nm{x'_i - x'_{i+1}} } \; , \label{eq:kernel_edge} 
\end{eqnarray}
where $\nm{.}$ denotes the Euclidean distance, the indices $i$ are taken modulo 3, and $K_{Feat}$ and $K_{Dist}$ are kernels functions introduced to compare pairs of labels from $\mathcal{L}$, and pairs of distances, respectively. It suffices now to define the kernels $K_{Feat}$ on $\mathcal{L} \times \mathcal{L}$ and $K_{Dist}$ on $\mathbb{R} \times \mathbb{R}$ in order to obtain, by (\ref{eq:kernel}), (\ref{eq:kernel_pharma}), (\ref{eq:kernel_node}) and (\ref{eq:kernel_edge}), a pharmacophore kernel for molecules. The first one compares the atom labels, while the second compares the distances between atoms in the pharmacophores. Intuitively they define the basic notions of similarity involved in the pharmacophore comparison, which in turns defines the overall similarity between molecules.

The kernel we use for $K_{Dist}$ is the Gaussian radial basis function (RBF) kernel,  known to be a safe default choice for SVM working on real numbers or vectors \cite{Scholkopf2002Learning}:
\begin{equation}\label{eq:rbf}
K_{Dist}^{RBF}(x,y) = \exp\left( -\frac{||x-y||^2 }{2 \sigma^2} \right)\;,
\end{equation}
where $\sigma>0$ is the bandwith parameter that will be optimized as part of the training of the classifier (see Section \ref{sec:experimental-setup}).

Concerning the kernel $K_{Feat}$ between labels, we investigate several choices. The labels belonging in principle to a finite set of possible labels, e.g., the set of atom types with their charges ($C$, $C^+$, $C^-$, N, ...), the following \emph{Dirac kernel} is a natural default choice to compare a pair of atom labels $l, l' \in \mathcal{L}$ : 
\begin{equation}\label{eq:categorical}
K_{Feat}^{Dirac}(l,l') = \begin{cases} 1 &\text{ if } l = l' \;,\\
                0 &\text{ otherwise}\;. 
                \end{cases}
\end{equation}
Alternatively, it might be relevant for pharmacophore definition to compare atoms not only on the basis of their types and partial charges, but also in terms of other physicochemical parameters such as their size, polarity and electronegativity. Formally, a physicochemical parameter for an atom with label $l$ is a real number $f(l)$. In that case, the Gaussian RBF kernel (\ref{eq:rbf}) could be applied directly to the parameter values to compare labels. A practical drawback of this kernel, however, is that it never vanishes. This induces an important computational burden compared to the Dirac kernel (see Section \ref{sec:computation}). As a result, we prefer to use the related \emph{triangular kernel}:
\begin{equation}\label{eq:triangle}
K_{Feat}^{Tri}(l,l')  = \begin{cases}  \frac{C - ||f(l)-f(l')||}{C} & \text{if }  ||f(l)-f(l')|| \leq C \;, \\ 
                            0 & \text{otherwise}\;.
                        \end{cases}
\end{equation}
An important difference between the Gaussian RBF and triangular kernels lies in the fact that the triangular kernel has a compact support, which means that it can be equal to $0$ for different atoms, resulting in important computational gain. The parameter $C$, to be optimized during the training phase of the algorithm, represents the range beyond which the kernel vanishes.

Note finally that the Gaussian RBF (\ref{eq:rbf}), triangular (\ref{eq:triangle}), and Dirac (\ref{eq:categorical}) kernels are known to be positive definite \cite{Scholkopf2002Learning}, and it follows from the closure properties of the family of kernel functions, that the kernel between pharmacophores $K_{\mathcal{P}}$ is valid for any choice of the kernels $K_{Dist}$ and $K_{Feat}$ proposed above.

\section{Kernel computation}\label{sec:computation}
We are now left with the task of computing the pharmacophore kernel (\ref{eq:kernel}) for a particular choice of feature and distance kernels $K_{Feat}$ and $K_{Dist}$. In this section we provide a simple analytical formula for this computation.

For any pair of molecules $m = \cbr{ \br{x_i,l_i } \in \mathbb{R}^3 \times \mathcal{L} }_{i=1,...,|m|}$ and $m' = \cbr{ \br{x'_i,l'_i } \in \mathbb{R}^3 \times \mathcal{L} }_{i=1,...,|m'|}$, let us define a square matrix $M$ of size $ n = |m|\times|m'|$, whose dimensions are indexed by the Cartesian product of $m$ and $m'$. In other words, to each index $i \in [1,n]$ corresponds a unique couple of indices $(i_1,i_2) \in [1,|m|] \times [1,|m'|]$, and to each dimension of the matrix $M$ corresponds a distinct pair of points taken from the molecules $m$ and $m'$. Denoting by $\one{.}$ the indicator function equal to one if its argument is true, zero otherwise, the entries of $M$ are defined by:
\begin{eqnarray}
M[i,j]  & = & M[(i_1,i_2),(j_1,j_2)] \notag\\ 
        & = &K_{Feat}\big( l_{i_1},l'_{i_2} \big) \times K_{Dist}\big( ||x_{i_1} - x_{j_1}|| , ||x'_{i_2} - x'_{j_2}|| \big) \times \one{i_1 \neq j_1} \times \one{i_2 \neq j_2} \; . \label{eq:matrixM}
\end{eqnarray}
The value of the pharmacophore kernel between $m$ and $m'$ can now be deduced from the matrix $M$ by the following result: 
\begin{proposition}\label{prop:kernel_computation}
The pharmacophore kernel (\ref{eq:kernel}) between a pair of molecules $m$ and $m'$ is equal to: 
$$
K(m,m') = \text{trace}(M^3) \; ,
$$
where $M$ is the square matrix of dimension $\abs{m} \times \abs{m'}$ constructed from $m$ and $m'$ by (\ref{eq:matrixM}).
\end{proposition}

\begin{proof}\label{proof:traceM3}

Developing the matrix products involved in the expression of $M^3$ we get 
$$
\text{trace}(M^3) = \sum_{i,j,k=1}^{n} M[i,j] M[j,k] M[k,i]\; ,
$$
where $n = |m|\times|m'|$ is the size of $M$. Using the fact that the indices of $M$ ranges over the Cartesian product of the set of indices $[1,|m|]$ and $[1,|m'|]$, we can rewrite this expression as :
\ignore{
\begin{eqnarray*}
\text{trace}(M^3) & = &  \sum_{i,j,k=1}^{n} M[i,j] M[j,k] M[k,j] \\
                  & = & \sum_{\substack{i=1,\\i=(i_1,i_2)}}^{n} \sum_{\substack{j=1,\\j=(j_1,j_2)}}^{n} \sum_{\substack{k=1,\\k=(k_1,k_2)}}^{n} M[(i_1,i_2),(j_1,j_2)] M[(j_1,j_2),(k_1,k_2)] M[(k_1,k_2),(i_1,i_2)] \\
                  & = & \sum_{i_1,j_1,k_1 = 1}^{|m|} \sum_{i_2,j_2,k_2 = 1}^{|m'|}  M[(i_1,i_2),(j_1,j_2)] M[(j_1,j_2),(k_1,k_2)] M[(k_1,k_2),(i_1,i_2)].
\end{eqnarray*}
}
$$
\text{trace}(M^3) = \sum_{i_1,j_1,k_1 = 1}^{|m|} \sum_{i_2,j_2,k_2 = 1}^{|m'|}  M[(i_1,i_2),(j_1,j_2)] M[(j_1,j_2),(k_1,k_2)] M[(k_1,k_2),(i_1,i_2)] \; .
$$
Substituting with the definition of $M$ given in (\ref{eq:matrixM}), we obtain :
\begin{eqnarray*}
\text{trace}(M^3) & = & \sum_{i_1,j_1,k_1 = 1}^{|m|} \sum_{i_2,j_2,k_2 = 1}^{|m'|}  \one{i_1 \neq j_1} \one{j_1 \neq k_1} \one{k_1 \neq i_1} \times \one{i_2 \neq j_2} \one{j_2 \neq k_2} \one{k_2 \neq i_2} \\
                  &  & \qquad \times    K_{Feat}\br{l_{i_1} , l'_{i_2}} \times K_{Dist}\br{||x_{j_1}-x_{i_1} || , ||x'_{j_2}-x'_{i_2}||} \\
                  &  & \qquad \times   K_{Feat}\br{l_{j_1} , l'_{j_2}} \times K_{Dist}\br{||x_{k_1}-x_{j_1} || , ||x'_{k_2}-x'_{j_2}||} \\
                  &  & \qquad \times   K_{Feat}\br{l_{k_1} , l'_{k_2}} \times K_{Dist}\br{||x_{i_1}-x_{k_1} || , ||x'_{i_2}-x'_{k_2}||} \\
                  & = & \sum_{i_1,j_1,k_1 = 1}^{|m|} \sum_{i_2,j_2,k_2 = 1}^{|m'|} \one{i_1 \neq j_i \neq k_1} \times \one{i_2 \neq j_2 \neq k_2} \\
                  &  & \qquad \times K_{\mathcal{P}} \br{ \br{ (x_{i_1},l_{i_1}),(x_{j_1},l_{j_1}),(x_{k_1},l_{k_1}) } , \br{ (x'_{i_2},l'_{i_2}),(x'_{j_2},l'_{j_2}),(x'_{k_2},l'_{k_2}) } } \\
                 & = &  \sum_{\substack{i_1,j_1,k_1 = 1, \\ i_1 \neq j_1 \neq k_1}}^{|m|} \sum_{\substack{i_2,j_2,k_2 = 1,\\i_2 \neq j_2 \neq k_2}}^{|m'|} K_{\mathcal{P}} \br{\br{ (x_{i_1},l_{i_1}),(x_{j_1},l_{j_1}),(x_{k_1},l_{k_1}) } , \br{ (x'_{i_2},l'_{i_2}),(x'_{j_2},l'_{j_2}),(x'_{k_2},l'_{k_2}) } } \\
                 & = & \sum_{ p \in \mathcal{P}(m)}  \sum_{ p' \in \mathcal{P}(m') } K_{\mathcal{P}} (p,p') \\
                 & = & K(m,m') \; .
\end{eqnarray*}
\end{proof}

If we neglect the cost of the addition and product operations, and let $u$  be that of evaluating the basis kernels $K_{Feat}$ and $K_{Dist}$, the complexity of the kernel between pharmacophores $K_{\mathcal{P}}$ is $6u$.
Since the cardinality of the set of pharmacophore $\mathcal{P}(m)$ of the molecule $m$ is $|m|^3$, the complexity of the direct computation of the pharmacophore kernel given in definition \ref{def:kernel} is $(|m|\times|m'|)^3 \times 6u$.
On the other hand, the computation given in Proposition~\ref{prop:kernel_computation} is a two step process : 
\begin{itemize}
\item first, initialization of the matrix $M$ :  each of the $(|m|\times|m'|)^2$ entries is initialized by the product of a kernel $K_{Feat}$ with a kernel $K_{Dist}$, for a complexity of $(|m|\times|m'|)^2 \times 2u$
\item second, computation of the trace of $M^3$, which has a complexity of $(|m|\times|m'|)^3$
\end{itemize}
The global complexity of the matrix-based computation of the kernel is therefore $(|m|\times|m'|)^3 + (|m|\times|m'|)^2 \times 2u $, or equivalently $(|m|\times|m'|)^3 \times (1 + 2u/(|m|\times|m'|))$.
In comparison with the direct approach, the matrix-based implementation proposed in Proposition~\ref{prop:kernel_computation} reduces the number of basis kernels $K_{Dist}$ and $K_{Feat}$ to be computed and is therefore more efficient.

In any case, the complexity of the pharmacophore kernel computation is therefore $\mathcal{O}\br{\br{ |m| \times |m'| }^3}$.
Even for relatively small molecules (of the order of 50 atoms), this complexity becomes in practice a serious issue when the size of the dataset increases to thousands or tens of thousands of molecules.
However, we can note from the definition given in (\ref{eq:matrixM}), that the lines of $M$ corresponding to pairs of points $\br{x,l} \in m$ and $\br{x',l'} \in m'$ for which  $K_{Feat}(l,l') = 0 $ are filled with zeros.
Based on this consideration, we observe that the cost of computing the kernel can be reduced by limiting the size of the matrix $M$, according to the following proposition.

\begin{proposition}\label{prop:reduction} 
If we let $M_2$ be the reduced version of a square matrix $M_1$, where the null lines and the corresponding columns are removed, then $\text{trace}(M_2^3) = \text{trace}(M_1^3)$.
\end{proposition}

\begin{proof}
Let $n_1$ (resp. $n_2$) be the size of $M_1$ (resp. $M_2$), and define $P$ (resp. $N$) as the subset of the set of indices $[1,n_1]$ that corresponds to the non-null (resp. null) lines of $M_1$.
By definition, we have
\begin{eqnarray}
\text{trace}(M_1^3)     & = &  \sum_{i=1}^{n_1} M_1^3[i,i] \notag\\
                        & = & \sum_{i,j,k=1}^{n_1} M_1[i,j]M_1[j,k]M_1[k,i]\;.\label{eq:M3}
\end{eqnarray}
Moreover, if $i \in N$, then $M_1[i,j] = 0 \; \forall j \in [1,n_1]$.
As a consequence, the term $M_1[i,j]M_1[j,k]M_1[k,i]$ in the summations over $i$, $j$, and $k$ in (\ref{eq:M3}) is zero as soon as at least one index $i$, $j$ or $k$ is in the set $N$. It follows that 
\begin{eqnarray*}
\text{trace}(M_1^3)     & = &  \sum_{i,j,k \in P}M_1[i,j]M_1[j,k]M_1[k,i] \\
                        & = &  \sum_{i,j,k = 1}^{n_2} M_2[i,j]M_2[j,k]M_2[k,i] \\
                        & = & \text{trace}(M_2)\;.
\end{eqnarray*}

\end{proof}

Proposition \ref{prop:reduction} implies that the Cartesian product of $m$ and $m'$ involved in the matrix $M$ defined in (\ref{eq:matrixM}) can be restricted to the pairs of points for which the label kernel $K_{Feat}$ is non-zero.
In the case of the Dirac kernel (\ref{eq:categorical}) for discrete labels, this boils down to introducing a dimension in $M$ for any pair of atoms having the same label. This result can have important consequences in practice. Consider for example the case where the atoms of the molecules $m$ and $m'$ are uniformly distributed in $k$ classes of atom labels.
In this case,  the size of the matrix $M$ is equal to $k\br{|m|/k \times|m'|/k} = |m|\times|m'|/k$.
The complexity of the kernel computation is therefore $\mathcal{O}\br{(|m|\times|m'|/k)^3} = \mathcal{O}\br{(1/k^3) (|m|\times|m'|)^3}$.
It is therefore reduced by a factor $k^3$ in comparison with the original implementation. More generally this shows that important gains in memory and computation can be expected when the set of labels is increased. Section \ref{sec:results} discusses such a case in more details when the partial charges of atoms are included or not in the labels. This also justifies why the triangular kernel (\ref{eq:triangle}), with its compact support, can lead to much faster implementations than the Gaussian RBF kernel (\ref{eq:rbf}) when applied to the comparison of physicochemical properties of atoms.
Finally, in a similar way, the kernel $K_{Dist}$ to compare distances can be set to a compactly supported kernel, such as the triangular kernel (\ref{eq:triangle}).
This has the effect of introducing sparsity in the matrix $M$, allowing the kernel computation to benefit from sparse matrix algorithms. This possibility was not further explored in this work.

\section{Relation with graph kernels}\label{sec:graph-kernels}

In this section we show that the pharmacophore kernel can be seen as an extension of the walk-count graph kernels \cite{Gartner2003graph} to the 3D representation of molecules. The walk-count graph kernel is based on the representation of a molecule $m$ as a labeled graph $m = (\mathcal{V}, \mathcal{E})$, defined by a set of  vertices $\mathcal{V}$, a set of edges $\mathcal{E} \subset \mathcal{V} \times \mathcal{V}$ connecting pairs of vertices, and a labeling function $l : \mathcal{V} \cup \mathcal{E} \rightarrow \mathcal{A}$, assigning a label $l(x)$ in an alphabet $\mathcal{A}$ to any vertex or edge $x$.
In the case of molecules, the set of vertices $\mathcal{V}$ corresponds to the atoms of the molecule, and the edges of the graph are usually defined as the covalent bonds between the atoms of the molecules \cite{Gartner2003graph,Kashima2004Kernels,Mahe2005Graph}. In order to extend this 2D representation to a graph structure capturing 3D information, we propose to introduce an edge between any pair of vertices of the graph. Molecules are therefore seen as complete, atom-based graphs.
If we now define a walk of length $n$ as a succession of $n+1$ connected vertices, it is easy to see that there is a one-to-one correspondence between the set of pharmacophores $\mathcal{P}(m)$ of a molecule $m$, and its  set of self-returning walks of length-three, that we call $\mathcal{W}_3(m)$.
We can therefore write the pharmacophore kernel (\ref{eq:kernel}) as a walk-based graph kernel :
$$
K(m,m') = \sum_{ p \in \mathcal{P}(m)}  \sum_{ p' \in \mathcal{P}(m') } K_{\mathcal{P}} (p,p') = \sum_{w \in \mathcal{W}_3(m)} \sum_{w' \in \mathcal{W}_3(m')} K_{Walk}(w,w') \; ,
$$
where $K_{Walk}(w,w') = K_P(p,p')$ for the pair of walks $(w,w')$ corresponding to the pair of pharmacophores $(p,p')$.
More precisely, consider a pair of pharmacophores  $p = \br{\br{x_{i},l_{i}}}_{i=1,2,3}$ and $p' = \br{\br{x'_{i},l'_{i}}}_{i=1,2,3}$, and a corresonding pair of walks $w = \br{w_1,w_2,w_3,w_1}$ and $w' = \br{w'_1,w'_2,w'_3,w'_1}$.
There is a direct equivalence between $K_{\mathcal{P}}$ and $K_{Walk}$ if we choose to label the vertices of the graphs by the atom labels involved in the pharmacophore characterization, and to label the edges by the Euclidian distance between the atoms they connect. 
Indeed, in this case we can write :  
\begin{eqnarray*}
 K_{\mathcal{P}} (p,p') & = &  \prod_{i=1}^3 K_{Feat} \br{ l_{i},l'_{i} }  K_{Dist} \br{ \nm{x_i - x_{i+1}} , \nm{x'_i - x'_{i+1}} } \\
                        & = &  \prod_{i=1}^3 K_{Feat} \br{l(w_i),l(w'_i)} K_{Dist} \br{ l\br{(w_i,w_{i+1})} , l\br{(w'_i,w'_{i+1})} } \\
                        & = & K_{Walk}(w,w') 
\end{eqnarray*}

A striking point of this kernel between walks is that it can be factorized along the edges of the walks : 
\begin{eqnarray}
K_{Walk}(w,w')  & = & \prod_{i=1}^3 K_{Feat} \br{l(w_i),l(w'_i)} K_{Dist} \br{l\br{(w_i,w_{i+1})} , l\br{(w'_i,w'_{i+1})} } \notag \\
                & = & \prod_{i=1}^3 K_{Step}\br{(w_i,w_{i+1}),(w'_i,w'_{i+1})} \label{eq:step-kernel}
\end{eqnarray}

The pharmacophore kernel therefore formulates as a walk-based graph kernel, with a walk kernel factorizing along the edges of the walks.
It follows from \cite{Kashima2004Kernels} that it can be computed by the formalism based on product-graphs and powers of the adjacency matrix proposed in \cite{Gartner2003graph}, if the adjacency matrix of the product-graph is weighted by the walk-step kernels $K_{Step}$ (\ref{eq:step-kernel}).
Consequently, the matrix $M$ defined in (\ref{eq:matrixM}) and upon which is based the kernel computation of Proposition \ref{prop:kernel_computation}, can be seen as a weighted adjacency matrix of a product-graph defined on complete, atom-based, molecular factor graphs.

\section{Fast approximation}\label{sec:fast-approximation}
As an alternative to the costly computation presented in Section \ref{sec:computation}, we introduce in this section several fast approximations to the pharmacophore kernel based on a discretization of the pharmacophore space.

Our definition of pharmacophores is based on the atoms 3D coordinates, but they can equivalently be characterized by the pairwise distances between atoms.
In order to define discrete pharmacophores, we restrict ourselves to discrete sets of atom labels (e.g., the set of atom types), and we discretize the range of distances between atoms into a predefined number of bins. Each distance is then mapped to the index of the bin it falls in, and a discrete pharmacophore is defined by a triplet of atom labels together with a triplet of bin indices.
More formally, if the distance range is discretized into $p$ bins, the set of discrete pharmacophores is a finite set defined as $\mathcal{T}_3 = \mathcal{L}^3 \times [1,p]^3$, where $\mathcal{L}$ is the set of atom labels. 
Consider the mappings $\phi_0^{3pt}$ and $\phi_1^{3pt}$ from the set of molecules to the set of discrete pharmacophores $\mathcal{T_3}$, defined for the molecule $m$ as :
\begin{itemize}
\item $\phi_0^{3pt}(m) = \br{\phi_{t,0}(m)}_{t \in \mathcal{T}_3}$, where $\phi_{t,0}(m)$ is the number of times the pharmacophore $t$ is found in the molecule $m$,
\item $\phi_1^{3pt}(m) = \br{\phi_{t,1}(m)}_{t \in \mathcal{T}_3}$, where $\phi_{t,1}(m)$ equals 1 if the pharmacophore $t$ is found in the molecule $m$, and 0 otherwise.
\end{itemize}

With these new definitions at hand we propose three discrete kernels, 

\begin{definition}[Three-points spectrum kernel]
For a pair of molecules $m$ and $m'$, we define the \emph{three-points spectrum kernel} $K_{Spec}^{3pt}$ as 
\begin{equation}\label{eq:kernel-spectrum-pharma}
K_{Spec}^{3pt}(m,m') = \langle \phi_{0}^{3pt}(m) , \phi_{0}^{3pt}(m') \rangle = \sum_{t \in \mathcal{T}_3} \phi_{t,0}(m)\phi_{t,0}(m')\;.
\end{equation}
\end{definition}
Note that if we define the mapping $d : \mathcal{P} \mapsto \mathcal{T}_3$, such that $d(p)$ is the discretized version of the pharmacophore $p \in \mathcal{P}$, we can explicitly write the three-points spectrum kernel as a particular pharmacophore kernel (\ref{eq:kernel}):
$$
K_{Spec}^{3pt}(m,m') = \sum_{p \in \mathcal{P}(m)} \sum_{p' \in \mathcal{P}(m')} \one{d(p) = d(p')} \;.
$$ 
This equation shows that this is a crude pharmacophore kernel, based on a kernel for pharmacophores that simply checks if two given pharmacophores have identical discretized versions or not.

\begin{definition}[Three-points binary kernel]
For a pair of molecules $m$ and $m'$, we define the \emph{three-points binary kernel} $K_{Bin}^{3pt}$ as 
\begin{equation}\label{eq:kernel-bin-pharma}
K_{Bin}^{3pt}(m,m') = \langle \phi_{1}^{3pt}(m) , \phi_{1}^{3pt}(m') \rangle = \sum_{t \in \mathcal{T}_3} \phi_{t,1}(m)\phi_{t,1}(m')\;.
\end{equation}
\end{definition}

\begin{definition}[Three-points Tanimoto kernel]\label{def:tanimoto-3pt}For a pair of molecules $m$ and $m'$, we define the \emph{three-points Tanimoto kernel} $K_{Tani}^{3pt}$ as 
\begin{equation}\label{eq:kernel-tani-pharma}
K_{Tani}^{3pt}(m,m') = \frac{K_{Bin}^{3pt}(m,m')}{K_{Bin}^{3pt}(m,m)+K_{Bin}^{3pt}(m',m')-K_{Bin}^{3pt}(m,m')}\;.
\end{equation}
\end{definition}
Note that the mapping $\phi_{1}^{3pt}(m)$ corresponds to a classical \emph{pharmacophore fingerprint} representation of the molecule $m$, where the bitstring is indexed by the pharmacophores of $\mathcal{T}_3$.
As a result, the three-points Tanimoto kernel is the equivalent of the Tanimoto coefficient for pharmacophore fingerprints, and constitutes a standard pharmacophore-based similarity measure \cite{Pickett1996Diversity,Saeh2005Lead,Brown1997information}.
Note that the dimensionality of the feature spaces associated to these kernels corresponds to the cardinality of $\mathcal{T}_3$, which is by definition $(np)^3$ for a label set $\mathcal{L}$ of cardinality $n$ and $p$ distance bins.\\

Finally, we consider additional ``two-points pharmacophore'' versions of the kernels (\ref{eq:kernel-spectrum-pharma}), (\ref{eq:kernel-bin-pharma})  and (\ref{eq:kernel-tani-pharma}), based on pairs, instead of triplets, of atoms \cite{Ralaivola2005Graph}.
Letting  $\mathcal{T}_2$ be the set of all possible two-points pharmacophores, that is, pairs of atom types together with the bin index of the edge connecting them, and $\phi_0^{2pt}(m) = \br{\phi_{t,0}(m)}_{t \in \mathcal{T}_2}$ and $\phi_1^{2pt}(m) = \br{\phi_{t,1}(m)}_{t \in \mathcal{T}_2}$ be the mappings of the molecule $m$ to $\mathcal{T}_2$, corresponding to $\phi_0^{3pt}(m)$ and $\phi_1^{3pt}(m)$, we define the three following kernels.

\begin{definition}[Two-points spectrum kernel]
For a pair of molecules $m$ and $m'$, we define the \emph{two-points spectrum kernel} $K_{Spec}^{2pt}$ as : 
\begin{equation}\label{eq:kernel-spectrum-pair}
K_{Spec}^{2pt}(m,m') = \langle \phi_{0}^{2pt}(m) , \phi_{0}^{2pt}(m') \rangle = \sum_{t \in \mathcal{T}_2} \phi_{t,0}(m)\phi_{t,0}(m') \;.
\end{equation}
\end{definition}

\begin{definition}[Two-points binary kernel]
For a pair of molecules $m$ and $m'$, we define the \emph{two-points binary kernel} $K_{Bin}^{2pt}$ as : 
\begin{equation}\label{eq:kernel-bin-pair}
K_{Bin}^{2pt}(m,m') = \langle \phi_{1}^{2pt}(m) , \phi_{1}^{2pt}(m') \rangle = \sum_{t \in \mathcal{T}_2} \phi_{t,1}(m)\phi_{t,1}(m') \;.
\end{equation}
\end{definition}

\begin{definition}[Two-points Tanimoto kernel]
For a pair of molecules $m$ and $m'$, we define the \emph{two-points Tanimoto kernel} $K_{Tani}^{2pt}$ as : 
\begin{equation}\label{eq:kernel-tani-pair}
K_{Tani}^{2pt}(m,m') = \frac{K_{Bin}^{2pt}(m,m')}{K_{Bin}^{2pt}(m,m)+K_{Bin}^{2pt}(m',m')-K_{Bin}^{2pt}(m,m')}\;.
\end{equation}
\end{definition}

The following proposition justifies the use of these fast kernels with SVM:
\begin{proposition}
The kernels (\ref{eq:kernel-spectrum-pharma}), (\ref{eq:kernel-bin-pharma}), (\ref{eq:kernel-tani-pharma}), (\ref{eq:kernel-spectrum-pair}), (\ref{eq:kernel-bin-pair}) and (\ref{eq:kernel-tani-pair}) are positive definite.
\end{proposition}

\begin{proof}
Kernels (\ref{eq:kernel-spectrum-pharma}), (\ref{eq:kernel-bin-pharma}), (\ref{eq:kernel-spectrum-pair}), (\ref{eq:kernel-bin-pair}) are directly expressed as dot-products, and are consequently positive definite.
Kernels (\ref{eq:kernel-tani-pharma}) and  (\ref{eq:kernel-tani-pair}) follow the definition of the Tanimoto kernel  which is known to positive definite \cite{Ralaivola2005Graph}.
\end{proof}

The kernels (\ref{eq:kernel-spectrum-pharma}), (\ref{eq:kernel-bin-pharma}), (\ref{eq:kernel-tani-pharma}), (\ref{eq:kernel-spectrum-pair}), (\ref{eq:kernel-bin-pair}) and (\ref{eq:kernel-tani-pair}) can be computed efficiently using an algorithm derived from that used in the implementation of spectrum string kernels \cite{Leslie2002spectrum}.
We describe this algorithm in the case of the three-points kernels (\ref{eq:kernel-spectrum-pharma}), (\ref{eq:kernel-bin-pharma})  and (\ref{eq:kernel-tani-pharma}), its extension to the two-points kernels  (\ref{eq:kernel-spectrum-pair}), (\ref{eq:kernel-bin-pair}) and (\ref{eq:kernel-tani-pair}) being straightfoward.
Following the notation of Section \ref{sec:graph-kernels}, we represent molecules by complete, atom-based labeled graphs, with the difference that the set of atom labels $\mathcal{L}$ defining the vertices labels is considered to be discrete (e.g., the atom types), and the edges are now labeled by the bin index of the corresponding inter-atomic distance.
We consider the problem of computing the Gram matrix $K$ associated to such a set of molecular graphs $\big \{G_i = (\mathcal{V}_{G_i}, \mathcal{E}_{G_i}) \big\}_{i=1,...,n}$ for the kernels (\ref{eq:kernel-spectrum-pharma}), (\ref{eq:kernel-bin-pharma}) and (\ref{eq:kernel-tani-pharma}).
The alphabet $\mathcal{A}$, involved in the graph labeling function $l$ of section \ref{sec:graph-kernels}, is defined as $\mathcal{A} = L_V \cup L_E$, where $L_V$ is the set of vertices labels, corresponding to the set of atom labels $\mathcal{L}$, and $L_E$ is the set of edges labels, corresponding to the set of distance bins indices.

The algorithm is based on the manipulation of sets of walk pointers within each graph, according to a tree transversal process.
If we let $n$ and $p$ be the cardinalities of $L_V$ and $L_E$ respectively, we define a rooted, depth-four tree structuring the space of pharmacophores $\mathcal{T}_3$ as follows : 
\begin{itemize}
\item the root node has $n$ sons, corresponding to the $n$ possible vertex labels
\item the depth-one and depth-two nodes have $n \times p$ sons, corresponding to the $n \times p$ possible pairs of edge and vertex labels
\item the depth-three nodes have $p$ sons, corresponding to the $p$ possible edges labels, a leaf node being implicitly associated the vertex label of its depth-one ancestor.
\end{itemize}
A path from the root to a leaf node therefore corresponds to a triplet of disctinct vertex labels, together with a triplet of distinct edge labels.
There is therefore a one-to-one correspondence between the leaf nodes and the pharmacophores of $\mathcal{T}_3$.
The principle of the algorithm is to recursively transverse this tree until each leaf node (i.e., each potential pharmacophore) is visited.
During this process, a set of walk pointers is maintained within each molecule.
The pointers are recursively updated such that the pointed walks correspond to the pharmacophores under construction in the tree-transversal process.
When reaching a leaf node, the pointed walks correspond to the occurences of a particular pharmacophore $t$ in the molecules.
The mappings $\phi_{t,0}(G_i)$ and $\phi_{t,1}(G_i)$ can therefore be computed for the molecular graphs $\{G_i\}_{i=1,...,n}$, and the kernel matrix can be updated.

A pseudo code of the algorithm is given in Algorithms \ref{algo:compute}, \ref{algo:init-pointers}, \ref{algo:extend-pointers} and \ref{algo:update-gram}. 
Algorithm \ref{algo:compute} is the main program in charge of the tree-transveral process, and Algorithms \ref{algo:init-pointers}, \ref{algo:extend-pointers} and \ref{algo:update-gram} are subroutines, introduced to initialize the walks pointers, extend the pointed walks, and update the Gram matrix respectively.
This pseudo-code relies on the abstract types \emph{Pointer} and \emph{Label}, to represent the walk pointers involved in the algorithm, and the generic vertices and edges labels, belonging to $L_V$ and $L_N$ respectively.
Formally, a \emph{Pointer} object consists of two graph vertices: a \emph{start} and \emph{current} vertex, representing the first and the current vertices of the pointed walk being extended.
To maintain walks pointers within each molecule, we introduce a matrix of pointers \emph{walkPointers = Pointer[][]} : this matrix is initially empty, and during the walk extension process, \emph{walkPointers[i][j]} corresponds to the $j$th pointer of the molecular graph $G_i$.
The stopping criterion of the recursion is controled by an integer variable \emph{depth} corresponding to the depth in the tree during the transversal process.
It is initialized to zero and incremented at each recursive call.
When \emph{depth} is three, a depth-three node was reached in the tree, which corresponds to pointers on length-two walks in the graphs.
In the subsequent recursive step, \emph{depth} is four, and the pointers are updated to ensure that the extended walks correspond to self-returning ones.
A leaf node is then reached and the recursion terminates, leading to an update of the Gram matrix.
Note however that the recursion is aborted whenever the set of walk pointers becomes empty for all graphs, since we only need to reach the leaf nodes corresponding to the pharmacophores truly present in the set of graphs.
The Gram matrix is updated according to the spectrum (\ref{eq:kernel-spectrum-pharma}), binary (\ref{eq:kernel-bin-pharma}) or Tanimoto (\ref{eq:kernel-tani-pharma}) definition of the kernel, and we introduce a $n \times n$ Gram matrix $K$, initialized to zero, together with a binary variable \emph{kernelType} that can take the values 'spectrum', 'binary' or 'Tanimoto'.

Computing the Gram matrix $K$ simply requires a call to the \emph{COMPUTE} function of Algorithm~\ref{algo:compute} with these initialized data : \emph{COMPUTE(walkPointers, depth, K, kernelType)}, for a specified \emph{kernelType}, and where the Pointer array $walkPointers$ is empty, $depth$ equals zero and the Gram matrix $K$ is filled with zeros.
Note however that in the case of the Tanimoto kernel type, this procedure computes the 'raw' kernel that actually corresponds to the binary kernel (\ref{eq:kernel-bin-pharma}). The matrix $K$ must be further normalized according to defintion \ref{def:tanimoto-3pt}, that is : $K[i][j]  = \frac{K[i][j]}{K[i][i] + K[j][j] - K[i][j]}$.\\

The cost of this algorithm depends on the number of leaf nodes visited, and is therefore bounded by the total number of leaves of the tree, that is $(np)^3$ if the number of distinct vertex labels is $n$ and the number of distance bins is $p$.
However, the maximum number of distinct pharmacophores that can be found in the molecule $m$ is $|m|^3$, and we do not need to exhaustively transverse the tree.
This means that to compute the kernel between the molecules $m$ and $m'$, at most $\min(|m|^3,|m'|^3)$ leaves, corresponding to the common pharmacophores of $m$ and $m'$, need to be visited.
The complexity of the algorithm is therefore $\mathcal{O}\br{ \min \br{(np)^3, \min \br{|m|^3,|m'|^3}} }$\footnote{Note however that in the case of the Tanimoto kernel (\ref{eq:kernel-tani-pharma}), the self kernels have to be computed, and the worst case complexity of the algorithm is $\mathcal{O}\br{ \min \br{(np)^3, |m|^3 + |m'|^3}}$.}.
For small molecules, the cost of the kernel will therefore depend on their number of atoms, while it will depend on the size of the discrete pharmacophores space for large molecules.

Note finally that although we omit the details, the previous algorithm and complexity analysis hold for the two-points versions of the kernels : the tree involved in the recursive transversal process is smaller (a depth-two tree, with $n^2p$ leaf nodes), and the complexity is reduced to $\mathcal{O}\br{\min\br{n^2p , \min(|m|^2,|m'|^2)}}$.

\begin{algorithm}{COMPUTE(Pointer[][] $walkPointers$,Integer $depth$, Float[][] $K$, String $kernelType$)}
\caption{main program}
\label{algo:compute}
\begin{algorithmic}
\STATE $depth = depth +1$
\IF{$depth = 1$}
        \FOR{$label \in L_V$}
        \STATE $walkPointers$ = initPointers($label$)
        \STATE compute($walkPointers,depth$)
        \ENDFOR
\ELSE 
        \FOR{$label_1 \in L_V$}
                \FOR{$label_2 \in L_E$}
                        \STATE $walkPointers$ = extendPointers($walkPointers, depth, label_1,label_2$)
                        \IF{$iwalkPointers \neq [][]$}
                                \IF{$depth = 4$}
                                        \STATE updateGram($walkPointers, K, kernelType$)
                                \ELSE
                                        \STATE compute($walkPointers,depth$)
                                \ENDIF
                        \ENDIF
                \ENDFOR
        \ENDFOR
\ENDIF
\end{algorithmic}
\end{algorithm}

\begin{algorithm}{INITPOINTERS(Label $label$)}
\caption{Sub-routine 1 : initialize walks pointers}
\label{algo:init-pointers}
\begin{algorithmic}
\STATE $walkPointers$ = Pointer[][]
\FOR{$i=1,...,n$}
        \FOR{$v \in \mathcal{V}_{G_i}$}
                \IF{$l(v) = label$}
                        \STATE $walkPointers[i]$.addPointer(start = $v$, current = $v$)
                \ENDIF
        \ENDFOR
\ENDFOR
\RETURN $walkPointers$
\end{algorithmic}
\end{algorithm}

\begin{algorithm}{EXTENDPOINTERS(Pointer[][] $walkPointers_{in}$, Integer $depth$, Label $label_1$, Label $label_2$)}
\caption{Sub-routine 2 : extend walks pointers}
\label{algo:extend-pointers}
\begin{algorithmic}
\STATE $walkPointers_{out}$ = Pointer[][]
\FOR{$i = 1, ..., n$}
        \FOR{$ptr \in walkPointers_{in}[i]$}
                \FOR{$(ptr.current, v) \in \mathcal{E}_{G_i}$}
                        \IF{$l(v) = label_1 \wedge l\big( (ptr.current,v)  \big) = label_2$}
                                \IF{ not( $depth = 4 \wedge v \neq ptr.start$ )}
                                        \STATE $walkPointers_{out}[i]$.addPointer(start = $ptr.start$, current = $v$)
                                \ENDIF
                        \ENDIF
                \ENDFOR
        \ENDFOR
\ENDFOR
\RETURN $walkPointers_{out}$
\end{algorithmic}
\end{algorithm}

\begin{algorithm}{UPDATEGRAM(Pointer[][] $walkPointers$, Float [][] $K$, String $kernelType$)}
\caption{Sub-routine 3 : update Gram matrix}
\label{algo:update-gram}
\begin{algorithmic}
\FOR{$i=1,...,n$}
        \FOR{$j=1,...,n$}
                \IF{$walkPointers[i] \neq  [] \wedge walkPointers[j] \neq []$}
                        \IF{$kernelType = $ 'spectrum'}
                                \STATE $update = walkPointers[i].size() \times walkPointers[j].size()$
                        \ELSE
                                \STATE $update = 1$
                        \ENDIF
                        \STATE $K[i][j] = K[i][j] + update$
                        \IF{$i \neq j$}
                                \STATE $K[j][i] = K[j][i] + update$
                        \ENDIF
                \ENDIF
        \ENDFOR
\ENDFOR
\end{algorithmic}
\end{algorithm}

\section{Experiments}\label{sec:experiments}

We now turn to the experimental section. The problem considered here consists in building predictive models to distinguish \emph{active} from \emph{inactive} molecules on several protein targets. This problem is naturally formulated as a supervised binary classification problem that can be solved by SVM.

\subsection{Datasets}

We tested the pharmacophore kernel on several datasets used in a recent SAR study \cite{Sutherland2003Spline-fitting}.
More precisely, we considered the following four publicly available datasets \footnote{Available as supporting information of the original study at \texttt{http://pubs.acs.org/journals/jcisd8/}}: 
\begin{itemize}
\item  the \emph{BZR} dataset, a set of 405 ligands for the benzodiazepine receptor,
\item the \emph{COX} dataset, a set of 467 cyclooxygenase-2 inhibitors,
\item the \emph{DHFR} dataset, a set of 756 inhibitors of dihydrofolate reductase,
\item the \emph{ER} dataset, a set of 1009 estrogen receptor ligands.
\end{itemize}
These datasets contain the 3D structures of the molecules, together with a quantitative measure of their ability to inhibit a biological mechanism. 
Datasets were filtered and split into training and test sets according to a particular data preparation scheme detailed in \cite{Sutherland2003Spline-fitting}.
Table \ref{tab:datasets} gathers basic informations about the datasets involved in the study.
\begin{table}
\begin{center}
\begin{tabular}{l|c|c|c|c}
& \multicolumn{2}{|c|}{TRAIN} &  \multicolumn{2}{|c}{TEST} \\
\cline{2-5}
&  Pos & Neg & Pos & Neg \\
\hline
BZR   & 94  & 87  & 63 & 62 \\
COX   & 87  & 91  & 61 & 64 \\
DHFR  & 84  & 149 & 42 & 118 \\
ER    & 110 & 156 & 70 & 110 \\
\end{tabular}
\end{center}
\caption{Basic informations about the datasets considered.\label{tab:datasets}}
\end{table}

\subsection{Experimental setup}\label{sec:experimental-setup}
We investigated in this study a simple labeling scheme to describe each atom (hydrogen atoms were systematically removed), and therefore the potential pharmacophores: the label of an atom is composed of its type (e.g., $C$, $O$, $N$...) and the sign of its partial charge ($+,-$ or $0$). Hence the set of labels can be expanded as $\mathcal{L} = \cbr{C^+,C^0,C^-,O^+,O^0,O^-,\ldots}$. The partial charges account for the contribution of each atom to the total charge of the molecule, and were computed with the QuacPAC software developed by OpenEye \footnote{\texttt{http://www.eyesopen.com/products/applications/quacpac.html}}. It is important to note that, contrary to the physicochemical properties of atoms, partial charges depend on the molecule and describe the spatial distribution of charges. Although the partial charges take continuous values, we simply kept their signs for the labeling as basic indicators of charges in the description of pharmacophores. We call \emph{categorical kernel} the kernel resulting from this labeling, where the kernel between labels $K_{Feat}$ is the Dirac kernel (\ref{eq:categorical}) and the kernel between distances $K_{Dist}$ is the Gaussian RBF kernel (\ref{eq:rbf}).

Alternatively, we tested several variants of this basic categorical kernel. First, we tested the effect of the partial charges by removing them from the labels, and keeping the same Dirac and Gaussian RBF kernels for the labels and distances, respectively. In this case the label of an atom reduces to its type. Second, we tested alternatives to the Dirac kernel between labels, by taking into account similarities between physicochemical properties of atoms with different labels. We considered the four following properties, taken from \cite{Emsley1998Elements} :  the \emph{Van der Waals radius}, which represents the radius of an imaginary sphere enclosing the atom, the \emph{covalent radius}, corresponding to half of the distance between two identical covalently bonded atomic nuclei, the \emph{first ionization energy}, the energy required to strip it of an electron from the atom, and the \emph{electronegativity}, a measure of the ability of an atom or molecule to attract electrons in the context of a chemical bond.
The Van der Waals and covalent radii account for the steric property of atoms, while the two latter properties encode their electrostatic behavior. In these cases, for computational reasons, a triangular kernel was used to compare different atoms with respect to these properties. Third, we tested the six fast approximations mentioned in Section \ref{sec:fast-approximation} with our original labeling scheme ($3$- and $2$-points spectrum, binary and Tanimoto kernels).

In addition, we tested the state-of-the-art Tanimoto kernel based on the 2D structure of molecules \cite{Ralaivola2005Graph} to evaluate the potential gain obtained by including 3D information. This kernel is defined as the Tanimoto coefficient between fingerprints indicating the presence or absence of all possible molecular fragments of length up to $8$ in the 2D structure of the molecule, where a fragment refers to a sequence of atoms connected by covalent bonds. We note that this fingerprint is similar to classical 2D-fingerprints such as the Daylight representation\footnote{\texttt{http://www.daylight.com/dayhtml/doc/theory/theory.toc.html}}, with the difference that our implementation does not require to fold the fingerprint into a small-size vector \cite{Mahe2005Graph}.

The different kernels were implemented in C++, and the SVM experiment was conducted with the freely available Python machine learning package PyML\footnote{Available at \texttt{http://pyml.sourceforge.net}}. For each experiment, all parameters of the kernel and the SVM were optimized over a grid of possible choices on the training set only, to maximize the mean area under the ROC curve (AUC) \cite{Fawcett2003ROC} over an internal 10-fold cross-validation. The results on the test set correspond to the performance of the SVM with the selected parameters only. The optimized parameters include the width $\sigma \in \cbr{0.1,1,10}$ (in angstroms) of the Gaussian RBF kernel used to compare distances, the soft-margin parameter of the SVM over the grid $\cbr{0.1,0.5,1,1.5,...,20}$, the number of bins used to discretize the distances for the fast approximations over the grid $\cbr{4,6,8,\ldots,30}$ and the cut-off parameter $C$ of the triangular kernel when physicochemical properties are used. This later parameter was chosen among $3$ values chosen such that $10$, $25$ or $50\%$ of all atom pairs in the training set have a non-zero value for the kernel. The larger $C$, the more atoms of distinct types are matched by the kernel, but the longer the kernel computation.

\subsection{Results}\label{sec:results}
Table \ref{tab:results} shows the results of classification for the different kernel variants.
Each line corresponds to a kernel, and reports several statistics : the accuracy (fraction of correctly classified compounds), sensitivity (fraction of positive compounds that were correctly classified), specificity (fraction of negative compounds that were correctly classified), and AUC.
The first line corresponds to the basic categorical kernel. The following five lines show the results of the five variants of the categorical kernel obtained by modifying the kernel between labels (Dirac kernel for labels without partial charges, and triangular kernel for $4$ physicochemical properties). The results obtained by the six fast approximations follow. Finally, we added the performance obtained by the state-of-the-art 2D Tanimoto kernel based on the 2D structure of the molecules and the best results reported in the reference publication \cite{Sutherland2003Spline-fitting}.

The results of parameters optimization on the training set often led to similar choices for different kernels. For example, the width of the Gaussian RBF kernel to compare distances was usually selected at $0.1$ angstrom, which corresponds to a very strong constraint on the pharmacophore matching. The cut-off parameter for the triangular kernel was usually chosen to allow $10$ or $25\%$ of matches between atoms on the training set. Finally, the number of bins selected by the fast approximations to discretize the distances was usually between 20 and 30 bins.

The results show that in general, the five variants of the categorical kernel obtained by modifying the kernel between labels (lines 2-6) lead to similar or slightly worse performances than the categorical kernel. Removing the partial charges from atom labels decreases the accuracy by $3$ to $5\%$ on all datasets except COX, confirming that the partial charge information is important for the definition of pharmacophores. The variants based on the four physicochemical properties of atoms lead to results globally similar to those obtained with atom type labels without partial charges, from which they are deduced. This shows that, in the context of this study, subtle pharmacophoric features based on physicochemical parameters instead of simply the type of atoms could not be detected.

The fast pharmacophore kernel obtained by applying a Dirac kernel to check when pairs of candidate pharmacophores fall in the same bin of the discretized space (3pt-spectrum) systematically degrades the accuracy of $1$ to $5\%$ over all four datasets compared to the categorical kernel. This suggests that the gain in computation time obtained by discretizing the space and computing a 3D-fingerprint-like representation of molecules has a cost in terms of accuracy of the final model. A particular limitation of the fingerprint-based method is that two pharmacophores could remain unmatched in they fall into two different bins, although they might be very similar but close to the bins boundaries. In the case of the pharmacophore kernel, such pairs of similar pharmacophores would always be matched.

Interestingly, however, performances competitive with the categorical kernel are obtained by the fast 3pt-binary and 3pt-Tanimoto kernels. On the BZR dataset, the 3pt-binary kernel even gives the best performance. Contrary to the 3pt-spectrum kernel, these kernels are not pharmacophore kernels in the sense of Definition \ref{def:kernel}; however they are based on the same representation as the 3pt-spectrum kernel, the only difference being the way to obtain the kernel value from the fingerprint description. Note finally that these two kernels give overall similar results.

We observe moreover that except for the COX dataset, the discrete kernels based on two-points pharmacophores lead to significantly worse results than their three-points counterparts.

For each dataset, the results obtained with the 2D-Tanimoto kernel are significantly worse than those of the categorical kernel, with a decrease ranging from $3$ to $7\%$ on the different datasets. This confirms the relevance of 3D information for drug activity prediction, that motivated this work. Finally we note that on all but the COX dataset, the categorical kernel outperforms the best results of \cite{Sutherland2003Spline-fitting}, confirming the competitiveness of our method compared to state-of-the-art methods.

Regarding the computational complexity of the different methods, Table \ref{tab:computation-times} shows the time required to compute the kernel matrices on the BZR training set for different kernels, on a desktop computer, equipped with a Pentium 4 - 3.6 GHz processor, and 1 GB RAM. 
In the discrete version, the distance range was split into 24 bins, and as expected, the kernels based on the discretization of the pharmacophore space are faster than their counterparts by a factor of $4$ to $35$, depending on the type of labels used (with or without the partial charge information). 
In the exact kernel computation, the effect of removing the partial charges from the labels is to induce more matches between atoms and therefore, as discussed in Section \ref{sec:computation}, to drastically slow the computation by a factor of $12$, consistent with the theoretical estimate that dividing the size of the label classes by $k$ increases the speed by a factor $k^3$.

\begin{table}
\begin{center}
\begin{tabular}{l|c|c}
& Exact & Discrete \\
\hline
With charges     & 20'   & 6' \\
Without charges  & 249'  & 7'\\
\end{tabular}
\end{center}
\caption{Computation times in minutes needed to compute the different kernel matrices on the BZR training set. 
The first column refers to the computation of the exact kernel (\ref{eq:kernel_pharma}), and the second one to the approximate kernels (\ref{eq:kernel-spectrum-pharma}) and (\ref{eq:kernel-tani-pharma}).}\label{tab:computation-times}
\end{table}

\section{Discussion and conclusion}\label{sec:discussion}
This paper presents an attempt to extend the application of recent machine learning algorithms for classification to the manipulation of 3D structures of molecules. This attempt is mainly motivated by applications in drug activity prediction, for which 3D pharmacophores are known to play important roles. Although previous attempts to define kernels for 3D structures (similar in fact to the 2pt-spectrum kernel we tested) led to mixed results \cite{Swamidass2005Kernels}, we obtained performance competitive with state-of-the-art algorithms for the categorical kernel based on the comparison of pharmacophores contained in the two molecules to be compared. This kernel is not an inner product between fingerprints, and therefore fully exploits the mathematical trick that allows SVM to manipulate measures of similarities rather than explicit vector representations of molecules, as opposed to other methods such as neural networks. We even observed that for the closest fingerprint-based approximation obtained by discretizing the space of possible pharmacophores (3pt-spectrum kernel), the performance significantly decreases. This highlights the benefits that can be gained from the use of kernels, which provide a satisfactory answer to the common issue of choosing a ``good'' discretization of the pharmacophore space to make fingerprints: once discretized, pharmacophores falling on different sides of bins edges do not match although they might be very close. We notice that approaches based on fuzzy fingerprints \cite{Horvath2003Neighborhood}, for example, aim at correcting this effect by matching pharmacophores based on different distance bins.

Although the best overall method in this study is the categorical kernel, it is interesting to notice that very competitive results are obtained by the binary and Tanimoto kernels applied to the discretized pharmacophore representation. Compared to the 3pt-spectrum, the better performance of the binary kernel suggests that the choice of the functional form of the kernel given a representation of molecules can play a critical role in terms of performance. 
Representing pharmacophores by indicators (bits) of their presence rather than by their precise counts can be interpreted as a trivial way to emphasize rare versus frequent pharmacophores in the kernels.
Alternatively, it might be possible for example, to adopt more flexible schemes to weight the pharmacophores depending on their probability of appearance in the molecules, and to modify in a similar way the functional form of the pharmacophore kernels (Definition \ref{def:kernel}) to improve performance over the categorical kernel. 

Concerning the practical use of our approach for screening of large datasets, Table \ref{tab:computation-times} shows that, even for the fastest variants, the approach based on kernel methods can be computationally demanding even for relatively small datasets. In practice, however, the time to train the SVM can be smaller than the times presented in Table \ref{tab:computation-times} because not all entries of the matrix are required. Speeding up SVM and kernel methods for large datasets is currently a topic of interest in the machine learning community, and applications in virtual screening on large databases of molecules will certainly benefit from the advances in this field.

Among the possible extensions to our work, a first direction would be to test and validate different definitions and labeling for the vertices of the pharmacophores. We limited ourselves to the simplest possible $3$-points pharmacophores based on single atoms annotated by their types and partial charges. The method could be improved by testing other schemes known to be relevant features as basic components of pharmacophores. Instead of single atoms, it is for example possible to consider groups of atoms forming functional units instead of single atoms to form pharmacophores. A second possible extension is to generalize this work to pharmacophores with more points, e.g., $4$ or $5$. Although several results will not remain valid in this case, such as the expression of the kernel as the trace of a matrix, this could lead to more accurate models in cases where the binding mechanism is well characterized by such pharmacophores. Finally, a third promising direction that is likely to be relevant for many real-world applications is to take into accounts different conformers of each molecule. Indeed, it is well-known that the biological activity to be predicted is often due to one out of several conformers for a given molecule, which suggests to represent a molecule not as a single 3D structure but as a set of structures. The kernel approach lends itself particularly well to this extension, thanks to the possibility to define kernels between sets of structures from a kernel between structures, just like we defined a kernel between sets of pharmacophores from a kernel between single pharmacophores.

\begin{sidewaystable}
\centering
\begin{tabular}{l|c|c|c|c||c|c|c|c||c|c|c|c||c|c|c|c}
& \multicolumn{4}{|c||}{BZR} &  \multicolumn{4}{|c||}{COX} &  \multicolumn{4}{|c||}{DHFR} &  \multicolumn{4}{|c}{ER} \\
\cline{2-17}
& Acc. & Sens. & Spec. & AUC   & Acc. & Sens. & Spec. & AUC   & Acc. & Sens. & Spec. & AUC   & Acc. & Sens. & Spec. & AUC \\
\hline
Categorical   & 76.4 & 74.0 & 78.9 & 82.1 &    69.8 & 69.8 & 69.8 & 75.1 &    \bf{81.9} & 63.3 & 88.8 & 84.8 &   \bf{79.8} & 72.0 & 84.7 & 86.8 \\
\hline 
Categorical, no partial charges  
              & 74.3 & 73.6 & 75.0 & 81.5 &    70.0 & 68.5 & 70.9 & 74.6 &    78.1 & 65.2 & 82.7 & 82.2 &   77.6 & 71.7 & 81.4 & 87.2 \\
Van der Waals & 74.6 & 74.0 & 75.3 & 81.7 &    67.1 & 67.2 & 67.0 & 72.1 &    77.8 & 68.6 & 81.1 & 83.1 &   77.7 & 72.5 & 80.9 & 87.1 \\
Covalent      & 74.3 & 73.6 & 75.0 & 80.8 &    70.0 & 68.5 & 71.2 & 74.3 &    76.9 & 68.8 & 79.7 & 82.6 &   77.8 & 72.0 & 81.5 & 87.3 \\ 
Ionization    & 74.1 & 73.8 & 74.7 & 81.8 &    68.9 & 68.4 & 69.4 & 74.0 &    77.9 & 66.7 & 82.0 & 82.8 &   77.6 & 72.3 & 81.5 & 87.1 \\ 
Electroneg.   & 74.9 & 74.4 & 75.3 & 81.6 &    70.2 & 67.9 & 72.5 & 73.4 &    77.9 & 67.4 & 81.6 & 83.0 &   78.2 & 70.4 & 83.1 & 87.7 \\ 
\hline
3pt-spectrum 
             & 75.4 & 74.4 & 76.3 & 81.3 &    67.0 & 64.4 & 69.5 & 75.9 &    76.9 & 70.9 & 79.0 & 81.9 &    78.6 & 78.3 & 78.8 & 87.4 \\  
3pt-binary
             & \bf{78.5} & 74.4 & 82.6 & 81.5 &    68.2 & 70.5 & 65.9 & 74.8 &    80.8 & 66.2 & 85.9 & 81.1 &    79.3 & 74.7 & 82.2 & 87.5 \\  
3pt-Tanimoto 
             & 78.3 & 74.6 & 82.1 & 84.7 &    68.0 & 68.0 & 68.0 & 74.2 &    81.6 & 69.8 & 85.6 & 83.1 &    79.0 & 67.9 & 86.4 & 88.7 \\  
2pt-spectrum
             & 71.4 & 61.3 & 81.6 & 80.3 &    68.9 & 70.2 & 67.7 & 74.7 &    67.7 & 67.4 & 67.9 & 72.3 &    78.7 & 75.9 & 80.4 & 84.5 \\
2pt-binary
             & 72.3 & 66.5 & 78.2 & 77.2 &    71.3 & 71.0 & 71.6 & 76.5 &    66.5 & 78.3 & 62.3 & 76.2 &    75.6 & 87..8 & 67.8 & 84.8 \\
2pt-Tanimoto 
             & 75.0 & 69.7 & 80.5 & 80.3 &    69.8 & 67.0 & 72.3 & 74.2 &    72.4 & 71.9 & 72.5 & 80.6 &    74.3 & 85.6 & 67.1 & 85.1 \\
\hline
2D-Tanimoto  & 71.2 & 71.9 & 70.5 & 80.8 &    63.0 & 67.5 & 58.6 & 69.8 &    76.9 & 73.8 & 78.0 & 83.0 &   77.1 & 69.3 & 82.1 & 83.6 \\ 
\hline
Sutherland    & 75.2 & 70.0 & 81.0 & XXX  &    \bf{73.6} & 75.0 & 72.0 & XXX &     71.9 & 74.0 & 71.0 & XXX &    78.9 & 77.0 & 80.0 & XXX\\    
\end{tabular}
\caption{Classification of the test sets, after model selection on the training set.}\label{tab:results}
\end{sidewaystable}

\end{document}

%% file: header.tex
\newcommand {\ignore}[1]{}               

\newcommand {\br}[1]{\left(#1\right)}

\newcommand {\cbr}[1]{\left\{#1 \right\}}
\newcommand {\nm}[1]{\Arrowvert\, #1 \,\Arrowvert}

\newcommand {\abs}[1]{\left\vert\, #1 \,\right\vert}

\newcommand {\mc}[1]{\mathcal{#1}}

\newcommand{\RR}{\mathbb{R}}    

\renewcommand{\k}{K}

